\documentclass[a4paper,fleqn,usenatbib]{mnras}

\usepackage{newtxtext,newtxmath}
\usepackage[T1]{fontenc}
\usepackage{ae,aecompl}
\usepackage{graphicx}	
\usepackage{enumitem}
\usepackage{rotating}

\newcommand{\xmm}{{\sc{XMM}}\emph{-Newton}}
\newcommand{\gc}{$\gamma$\,Cas}
\newcommand{\hd}{HD\,119682}
\newcommand{\vc}{V767\,Cen}
\newcommand{\kms}{km\,s$^{-1}$}
\newcommand{\ha}{H$\alpha$}

\title[Disk changes in 2 \gc ]{X-ray response to disk evolution in two \gc\ stars\thanks{Based on \xmm, {\it Swift}, and ESO data}. }

\author[Y. Naz\'e et al.]{Ya\"el~Naz\'e$^1$\thanks{F.R.S.-FNRS Senior Research Associate, email: ynaze@uliege.be}, Gregor Rauw$^1$, Terrence Bohlsen$^2$, Bernard Heathcote$^3$, Padric Mc Gee$^4$, Paulo Cacella$^5$, \newauthor{and Christian Motch$^6$}
\\
$^1$ Groupe d'Astrophysique des Hautes Energies, STAR, Universit\'e de Li\`ege, Quartier Agora (B5c, Institut d'Astrophysique et de G\'eophysique), \\
All\'ee du 6 Ao\^ut 19c, B-4000 Sart Tilman, Li\`ege, Belgium\\
$^2$ SASER/Mirranook Observatory, Armidale NSW, Australia\\
$^3$ SASER/269 Domain Road, South Yarra, VIC, 3141, Australia\\
$^4$ SASER/Department of Physics, University of Adelaide, Adelaide 5005, Australia\\
$^5$ DogsHeaven Observatory, WMP Q25 CJ1 LT10 Brasilia DF, Brazil\\
$^6$ Universit\'e de Strasbourg, CNRS, Observatoire Astronomique de Strasbourg, UMR 7550, F-67000 Strasbourg, France
}

\pubyear{2021}

\begin{document}
\label{firstpage}
\pagerange{\pageref{firstpage}--\pageref{lastpage}}
\maketitle

\begin{abstract}
The H$\alpha$ emission of a set of southern \gc\ stars was monitored since 2019, with the aim of detecting transition events and examining how their peculiar X-ray emission would react in such cases. Two stars, \hd\ and \vc, were found to display slowly decreasing disk emissions. These decreases were not perfectly monotonic and several temporary and limited re-building events were observed. For \hd, the emission component in H$\alpha$ disappeared in mid-July 2020. In X-rays, the X-ray flux was twice smaller than recorded two decades ago but of a similar level as observed a decade ago. The X-ray flux decreased over the campaign by 30\%, but the hardness remained similar in datasets of all epochs. In particular, the \gc\ character remained as clear as before even when there was no trace of disk emission in the H$\alpha$ line. For \vc, the full disappearance of disk emission in H$\alpha$ never occurred. We followed closely a disk rebuilding event, but no significant change in flux or hardness was detected. These behaviours are compared to those of other \gc\ stars and their consequences on the X-ray generation are discussed.  
\end{abstract}


\section{Introduction}
The vast majority of stars are known X-ray emitters, and the most massive ones are no exception. The cause of their high-energy emission lies in the material ejected by those stars. The stellar winds of OB stars are driven by the scattering of the UV radiation by metallic ions. This driving process is intrinsically unstable and the resulting shocks lead to the generation of X-rays \citep{fel97}. This wind emission is soft (plasma temperature of about 0.6\,keV) and rather faint ($\log(L_{\rm X}/L_{\rm BOL}\sim -7$). If a large-scale magnetic field is present, the wind flows can be channeled by the field and collide, generating hot plasma. This leads to an additional X-ray emission, generally harder in character \citep{udd16}. Alternatively, when two massive stars form a binary system, the two winds may collide and the strong shock may generate X-rays \citep{rau16}: some massive binary systems thus appear harder (plasma temperature of about 2.\,keV) and brighter ($\log(L_{\rm X}/L_{\rm BOL})$ up to $-6$). Finally, a small subgroup of stars have emerged in the last decade: the \gc\ analogs \citep{smi16}. These stars also display thermal X-ray spectra, but much harder (plasma temperature $>5$\,keV) and brighter ($\log(L_{\rm X}/L_{\rm BOL})$ between $-6.2$ and $-4$). Only X-ray binaries appear brighter in the X-ray range than those \gc\ stars. It must be noted that all of these objects are of the Oe/Be spectral type, i.e. they possess a decretion disk. However, up to now, the only difference spotted between the other Oe/Be stars and the \gc\ analogs resides in their X-ray properties \citep{naz20tess,naz21}.

While the link between the outflows and the X-ray emission is well understood for the other cases (embedded wind shocks, magnetically confined winds, colliding winds), the origin of the peculiar \gc\ characteristics remains debated. Up to now, all detected \gc\ analogs belong to the Oe/Be category, i.e. they possess a decretion disk in Keplerian rotation. This gives us an important clue, but the exact role of the disk in the generation of the high-energy emission remains unclear. In this context, two broad classes of scenarios to explain the \gc\ phenomenon have been proposed. The first one relies on the presence of a companion. It could be a compact object (WD or NS) and the X-rays would then come from its accretion of material from the Be star and its disk \citep{mur86,pos17}. In such a case, if the disk dissipates, the source of material disappears hence the X-ray emission should ultimately (i.e. after some travelling delay) stop. Another possibility is to consider a stripped He-star companion whose stellar wind collides with the peripheral regions of the disk, thereby leading to a possible emission of X-rays \citep{lan20}. Again, disappearance of the disk implies cutting the feeding of the X-ray source. In contrast, the second type of scenario requires no companion as it involves magnetic interactions between the Be star and its inner disk \citep{rob02,mot15}. In such a case, if the disk fully dissipates, the interactions will no longer take place hence the peculiar X-ray emission should disappear, with a faster reaction time than for the previous case. Similar arguments can be built perforce if considering a disk outburst rather than a disk dissipation. Following the behaviour of the X-ray emission in reaction to changes in the disk therefore has an important diagnostic value. In this context, long-term monitorings are required as they are the only way to assess the amplitude of the correlated X-ray response to an optical event (if any) and to derive the time lag between them (if any). 

In this paper, we present the results of such an exercise for two \gc\ stars, \hd\ and \vc. The \gc\ nature of \hd\ (B0Ve, $V$=7.9) was first reported by \citet{rak06} and \citet{saf07}. Indeed, the X-ray emission is both much harder (plasma temperature of $\sim$10\,keV, hard-to-soft flux ratio of 2.5) and brighter ($\log(L_{\rm X}/L_{\rm BOL})\sim -5.7$) than for ordinary OB stars \citep{naz18}. The star was also classified as a binary candidate in \citet{naz21}. The X-ray emission of \vc\ (HD\,120991, B2Ve, $V$=6.1) was studied in detail by \citet{naz18} thanks to an \xmm\ archival exposure. The star displayed a $\log(L_{\rm X}/L_{\rm BOL})\sim -5.4$, a plasma temperature of 6.4\,keV, and a hard-to-soft flux ratio of 2, all pointing to a \gc\ nature. Unfortunately, no optical data were available for the two stars at the time of these discovery observations. However, our optical monitoring fills this gap and enlarges the view by revealing changes in their disks. X-ray observations were then triggered to assess the reaction of the high-energy emission to the disk variations. Section 2 presents the data used in this study, while Section 3 derives the results, Section 4 discusses them and Section 5 summarizes them. 

\section{Data}

\subsection{Optical data}
To ensure a regular monitoring of the \ha\ line, we set up a collaboration with amateur astronomers. Since 2019, four persons regularly observed a set of southern Be stars, with three observers (co-authors TB, BH, PMcG) based in Australia and one (co-author PC) in Brazil. Their instruments were 11--14 inches reflectors equipped with spectrographs (Gerlach LowSpec, Shelyak LHIRESIII, LISA, eShel) which provided spectral resolutions between 400 and 20\,000 (most common value was $\sim$4000 for \hd\ and $\sim$10000 for the brighter \vc). For \hd, exposure times ranged from 5 minutes to 3.5\,hrs, leading to typical signal-to-noise ratios of 75; for \vc, exposure times were similar but average signal-to-noise was 50. The spectra were reduced in a standard way using ISIS\footnote{http://www.astrosurf.com/buil/isis-software.html} and finally normalized over the same set of continuum windows using polynomials of low order. No telluric correction could be made for the low resolution spectra (all \hd\ data and a quarter of \vc\ data) but it was applied for observations taken with higher resolution. All Australian amateur spectra were deposited on the Be Star Spectra (BeSS) open-access database\footnote{http://basebe.obspm.fr}. Note that \hd\ belongs to a small group made of several stars, notably CPD--62$^{\circ}$3559 (K\,2\,II/III) and HD\,119699 (A\,1\,II), but their spectral types being different from that of \hd\ and their separation being large enough, the amateur spectra of \hd\ were not contaminated by these close neighbours.

In parallel, a few spectra of both stars were obtained at the Cerro Paranal ESO Observatory for our ESO program ID 105.204D. They were taken with the Ultraviolet and Visual Echelle Spectrograph (UVES) in dichroic mode (covered regions: 3300--4560\,\AA\ at $R\sim 70\,000$ and 4730--6830\,\AA\ at $R\sim 100\,000$). These spectra were already used and presented in \citet{naz21}, and we refer to that publication for details on this dataset. An additional X-Shooter spectrum of each star was also taken for the same program. Their spectral resolution is lower than that of UVES ($R\sim 20\,000$ in the visible range) but the signal-to-noise ratios are high ($\sim200$). These spectra were reduced in the same way as the UVES data.

H$\alpha$ equivalent widths were estimated for all spectra using the first-order moment over a given velocity range (--600 to 600\,\kms). They are listed in Tables \ref{ew} and \ref{ew2} for \hd\ and \vc, respectively.

\begin{table}
  \caption{Equivalent widths measured on H$\alpha$ (in the --600 to +600\,\kms\ interval) for \hd.  \label{ew}}
\setlength{\tabcolsep}{3.3pt}
  \begin{tabular}{lcc|lcc}
    \hline
Date & ID & $EW$ (\AA) & Date & ID & $EW$ (\AA) \\
        \hline
8564.993 &BH$^n$  &   -4.29$\pm$0.12 & 9049.936 &TB      &    1.25$\pm$0.02 \\ 
8573.977 &TB      &   -3.81$\pm$0.03 & 9053.961 &PMcG$^l$&    1.12$\pm$0.02 \\ 
8580.956 &TB      &   -3.53$\pm$0.03 & 9057.953 &PMcG$^l$&    1.22$\pm$0.03 \\ 
8602.994 &BH$^n$  &   -3.04$\pm$0.05 & 9060.885 &BH$^l$  &    1.12$\pm$0.04 \\ 
8610.011 &BH$^n$  &   -3.12$\pm$0.11 & 9061.954 &TB      &    1.21$\pm$0.03 \\ 
8661.063 &PMcG$^l$&    0.07$\pm$0.07: & 9074.952 &TB$^l$  &    0.80$\pm$0.03 \\ 
8662.012 &TB$^l$  &   -1.03$\pm$0.06 & 9080.974 &TB      &    0.59$\pm$0.04 \\ 
8676.913 &TB      &   -0.19$\pm$0.02 & 9089.902 &BH$^l$  &    0.00$\pm$0.07 \\ 
8682.954 &TB      &    0.14$\pm$0.02 & 9098.880 &BH$^l$  &    0.34$\pm$0.07 \\ 
8698.991 &TB      &    0.30$\pm$0.03 & 9185.837 &UVES    &   1.650$\pm$0.003 \\
8713.963 &TB      &    1.06$\pm$0.03 & 9205.842 &UVES    &   1.639$\pm$0.003 \\
8719.949 &TB      &    1.15$\pm$0.02 & 9218.850 &Xsh     &   1.007$\pm$0.012 \\
8721.940 &TB      &    0.97$\pm$0.02 & 9234.113 &PMcG$^l$&    1.39$\pm$0.04  \\
8724.925 &PMcG    &    0.21$\pm$0.06 & 9242.753 &UVES    &   1.186$\pm$0.003 \\
8728.935 &TB      &    0.99$\pm$0.03 & 9256.984 &TB      &    1.25$\pm$0.05  \\
8756.956 &PMcG$^n$&    1.42$\pm$0.27 & 9257.801 &UVES    &   1.226$\pm$0.004 \\ 
8837.225 &PMcG$^n$&   -3.06$\pm$0.14 & 9274.662 &UVES    &  -0.294$\pm$0.003 \\ 
8865.210 &PMcG$^n$&   -2.68$\pm$0.13 & 9275.999 &TB      &   -0.53$\pm$0.03  \\
8878.205 &PMcG    &   -1.98$\pm$0.06 & 9277.079 &PMcG$^l$&   -0.27$\pm$0.04  \\
8934.002 &PMcG$^n$&    1.92$\pm$0.17 & 9297.953 &TB      &   -0.88$\pm$0.03  \\
8952.069 &TB      &    1.06$\pm$0.03 & 9329.994 &TB      &   -0.74$\pm$0.02  \\
8974.035 &PMcG$^l$&    1.36$\pm$0.07 & 9331.188 &PMcG$^l$&   -0.60$\pm$0.05  \\
8981.055 &TB      &    1.75$\pm$0.02 & 9364.989 &BH$^l$  &    0.98$\pm$0.03  \\
9017.003 &TB      &    1.57$\pm$0.02 & 9371.045 &TB      &    1.07$\pm$0.02  \\
9037.950 &TB      &    1.51$\pm$0.03 & 9394.925 &PMcG$^l$&    1.07$\pm$0.03  \\
9044.996 &TB      &    1.48$\pm$0.03 & 9424.977 &TB      &    0.77$\pm$0.03  \\
9045.021 &PMcG$^l$&    1.36$\pm$0.04 \\
\hline
  \end{tabular}
  
{\scriptsize Dates are in the format HJD-2\,450\,000, and ':' indicates an uncertain value. The ID refers to the source of the spectrum: UVES or Xsh for ESO data, TB, BH, PMcG or PC for the amateur data (from their initials, see authors' list). Symbols $^n$ and $^l$ are added for the noisier spectra ($SNR<20$) and lower resolution data ($R<3000$ but $SNR>20$), respectively. Note that the errors on equivalent widths are computed from flux errors: they only reflect the $SNR$ and do not include the normalization errors. }
\end{table}

\begin{table}
  \caption{Same as in Table \ref{ew} for \vc, except that low resolution is here defined as $R<5000$.  \label{ew2}}
\setlength{\tabcolsep}{3.3pt}
  \begin{tabular}{lcc|lcc}
    \hline
Date & ID & $EW$ (\AA) & Date & ID & $EW$ (\AA)\\
        \hline
8570.959 & TB     & -8.25$\pm$0.01   & 9116.885 & TB     & -3.55$\pm$0.01   \\
8578.070 & BH     & -7.46$\pm$0.03   & 9127.873 & TB     & -2.15$\pm$0.03   \\ 
8590.038 & BH     & -7.09$\pm$0.02   & 9191.842 & UVES   & -3.981$\pm$0.002 \\ 
8599.002 & TB     & -7.15$\pm$0.01   & 9196.843 & Xsh    & -3.889$\pm$0.013 \\ 
8610.154 &PMcG$^l$& -8.94$\pm$0.03   & 9206.834 & UVES   & -3.436$\pm$0.003 \\ 
8624.000 & BH     & -10.29$\pm$0.02  & 9231.859 & UVES   & -2.502$\pm$0.003 \\ 
8641.155 &PMcG$^l$& -7.68$\pm$0.05   & 9234.149 &PMcG$^l$& -2.64$\pm$0.03   \\ 
8655.954 &PMcG$^l$& -8.07$\pm$0.08   & 9246.777 & UVES   & -1.864$\pm$0.002 \\ 
8668.932 & BH     & -7.88$\pm$0.02   & 9298.965 & TB     & -2.24$\pm$0.01   \\ 
8683.984 &PMcG$^l$& -7.39$\pm$0.02   & 9335.978 & TB     & -3.42$\pm$0.01   \\ 
8684.995 & TB     & -7.30$\pm$0.01   & 9351.943 & TB     & -2.31$\pm$0.01   \\ 
8685.943 & BH     & -7.31$\pm$0.03   & 9364.964 & BH$^l$ & -0.96$\pm$0.03   \\ 
8693.970 & TB     & -7.31$\pm$0.01   & 9368.914 & BH     & -0.91$\pm$0.04   \\ 
8697.971 & TB     & -7.76$\pm$0.02   & 9370.923 & TB     & -1.62$\pm$0.01   \\ 
8724.949 &PMcG$^l$& -6.85$\pm$0.03   & 9381.914 & BH     & -0.62$\pm$0.03   \\ 
8728.953 & TB$^l$ & -5.99$\pm$0.03   & 9386.933 & BH     & -1.05$\pm$0.03   \\ 
8865.177 &PMcG$^l$& -5.01$\pm$0.05   & 9394.989 &PMcG$^l$& -0.33$\pm$0.04   \\ 
8934.037 &PMcG$^l$& -2.15$\pm$0.04   & 9398.501 & PC     & -0.13$\pm$0.04   \\ 
8937.978 & TB     & -2.42$\pm$0.02   & 9399.534 & PC     & -0.36$\pm$0.03   \\ 
8948.992 & BH     & -2.56$\pm$0.05   & 9401.876 &PMcG$^l$& -0.91$\pm$0.03   \\ 
8953.044 &PMcG$^l$& -2.85$\pm$0.03   & 9403.964 & BH     & -0.78$\pm$0.03   \\ 
8960.996 & BH     & -3.41$\pm$0.02   & 9406.467 & PC     & -1.33$\pm$0.02   \\ 
8962.024 & TB     & -3.42$\pm$0.01   & 9406.977 & TB     & -2.00$\pm$0.01   \\ 
8972.955 & TB     & -3.10$\pm$0.02   & 9407.505 & PC     & -1.50$\pm$0.03   \\ 
8974.080 &PMcG$^l$& -3.34$\pm$0.03   & 9408.465 & PC     & -2.17$\pm$0.02   \\ 
8975.994 & TB     & -2.92$\pm$0.02   & 9410.462 & PC     & -1.92$\pm$0.03   \\ 
8984.037 & BH     & -3.68$\pm$0.03   & 9421.462 & PC     & -4.08$\pm$0.03   \\ 
8995.951 & BH     & -2.92$\pm$0.03   & 9422.959 & TB     & -4.62$\pm$0.01   \\ 
8998.034 & TB     & -2.70$\pm$0.02   & 9424.973 & BH$^l$ & -4.05$\pm$0.02   \\ 
9003.996 & TB     & -2.46$\pm$0.02   & 9426.474 & PC     & -3.76$\pm$0.03   \\ 
9015.943 & TB     & -1.61$\pm$0.02   & 9428.909 & BH$^l$ & -3.66$\pm$0.01   \\ 
9018.024 & BH     & -1.35$\pm$0.04   & 9429.473 & PC     & -4.06$\pm$0.03   \\ 
9029.970 & BH     & -2.11$\pm$0.03   & 9435.442 & PC     & -4.39$\pm$0.03   \\ 
9031.971 & TB     & -2.19$\pm$0.02   & 9437.976 & TB     & -4.55$\pm$0.01   \\ 
9044.996 & BH     & -3.28$\pm$0.02   & 9441.447 & PC     & -4.07$\pm$0.02   \\ 
9045.074 &PMcG$^l$& -3.26$\pm$0.02   & 9445.908 & BH     & -3.42$\pm$0.03   \\ 
9062.973 & BH     & -3.33$\pm$0.04   & 9446.994 & TB     & -3.81$\pm$0.02   \\ 
9066.997 & TB     & -3.09$\pm$0.02   & 9456.907 & BH     & -2.99$\pm$0.04   \\ 
9071.891 & BH$^l$ & -2.46$\pm$0.03   & 9459.969 & TB     & -3.49$\pm$0.02   \\ 
9074.983 & TB     & -2.80$\pm$0.02   & 9466.915 & TB     & -2.99$\pm$0.02   \\ 
9106.910 & BH$^n$ & -1.56$\pm$0.06   & 9481.916 & TB     & -1.97$\pm$0.02   \\
9111.496 & UVES   & -2.312$\pm$0.002 & 9591.165 & PMcG   & -1.50$\pm$0.02   \\
9112.914 & BH$^n$ & -3.32$\pm$0.07   &  \\
\hline
  \end{tabular}
  \end{table}

Contemporaneous photometry was recorded by ASAS-SN\footnote{https://asas-sn.osu.edu/} for both stars. The targets are however brighter than the saturation limit of ASAS-SN. In such cases, a correction procedure using the bleed trails is implemented, but it is far from perfect and notably requires the targets to be isolated (ASAS-SN has a PSF of 15\arcsec\ FWHM). For \hd, which lies in a crowded area, the data points are unrealistically spread over several magnitudes and the lightcurve is unusable. For \vc, the situation appears better, although a dispersion of $\sim 0.07$\,mag is present (which cannot hide long-term trends if present, see e.g. a similar situation for $\pi$\,Aqr in \citealt{naz19piaqr}). Only g-band data cover the dates of our monitoring. A few outliers (having values deviating from the median by 3 times the median absolute deviations) were filtered out.

\subsection{X-ray observations}
Both our targets were observed with \xmm. The first observation of \hd\ was taken in August 2001 in full frame mode and with a thick filter to avoid contamination by UV/optical photons (40\,ks, Rev.\,315, ObsID=0087940201, PI Hughes). \hd\ here appears off-axis as the observation was concerned with a nearby supernova remnant. The star was re-observed on-axis in March 2009 (54\,ks, Rev.\,1692, ObsID=0551000201, PI Motch), this time with a medium filter. Finally, we triggered our TOO program to monitor the X-ray emission of \hd\ as its disk was disappearing, leading to four additional observations taken with the medium filter and in Large window mode (PI Rauw; 10\,ks in August 2019 on Rev.\,3610, ObsID=0840310901; 20\,ks in January 2020 on Rev.\,3684, OBsID=0840311001; 10\,ks in July 2020 for Rev.\,3775, ObsID=0840311101 and March 2021 for Rev.\,3890, ObsID=0840310801).

\xmm\ also observed \vc\ during 7\,ks in January 2007 (Rev. 1306, ObsID=0402121801, PI Favata) using a thick filter and the large window mode. Following the observation of a disk flaring (see below), we requested a TOO observation which was taken in July 2021 with the same characteristics as in the first pointing (Rev. 3967, ObsID=0891800801, PI Naz\'e).

All \xmm\ data were processed with the Science Analysis Software (SAS) v19.1.0 using calibration files available in June 2021 and following the recommendations of the \xmm\ team\footnote{SAS threads, see \\ http://xmm.esac.esa.int/sas/current/documentation/threads/ }. The European Photon Imaging Camera (EPIC) observations were first processed with the pipeline and then filtered to keep only the best-quality data ({\sc{pattern}} 0--12 for MOS and 0--4 for pn). To assess whether contamination by background proton flares were present, we built global light curves for energies above 10\,keV and discarded time intervals corresponding to flares. Only the older two datasets and that of July 2020 were affected by such flares for \hd\ while there was no flare for \vc\ data. Source detection was then performed to assess the crowding in the fields-of-view. This allowed us to carefully choose extraction regions. The source regions were circles centered on the Simbad positions of the targets and with radii of 30\arcsec\ in general while background regions were chosen from nearby circles devoid of sources and generally 50\arcsec\ in radius. Background-corrected lightcurves were calculated for energy bands of 0.5--10.\,keV, 0.5--2.\,keV, and 2.--10.\,keV. For \hd, bins of 100\,s and 1\,ks were used, whereas shorter bins of 50\,s and 500\,s could be used for \vc\ to get the same lightcurve quality since its X-ray flux is larger. The lightcurves were corrected for vignetting, off-axis angle, and bad pixels, and bins exposed during less than half nominal bin length were eliminated. For spectra, dedicated calibration matrices were built and a grouping was then applied to obtain an oversampling factor of maximum five and a minimum signal-to-noise ratio of 3. 

For \hd, a few {\it Chandra} observations were also available. These grating observations were taken in December 2008 (ObsID=8929 and 10834--6, PI Rakowski), totalling nearly 150\,ks altogether. Individual zeroth order spectra as well as combined (order +1 and --1) grating spectra were extracted for HEG and MEG. The reduction of these observations was presented in \citet{naz18} and no further processing was applied for this paper. As mentioned in \citet{naz18}, additional {\it Chandra} observations of \hd\ were taken with ACIS-I and suffer from pile-up hence could not be used.

For \vc, we further obtained an X-ray monitoring with the Neil Gehrels {\it Swift} observatory (ObsID=00014422001--8) during the second semester of 2021 and early 2022. The X-ray telescope (XRT) was used in Windowed Timing mode because \vc, with its $V$=6.1, is too bright in the optical and UV ranges for the other observation mode. Since we were mostly interested in flux variations, exposures were of 2--3\,ks duration, allowing for a few hundred counts to be collected for the source. Note that exposures 00014422003--4, taken four days apart, were combined to reach the required signal-to-noise. Individual count rates and spectra were obtained with the {\it Swift} on-line tool\footnote{https://www.swift.ac.uk/user\_objects/}. Note that \vc\ is at the limit for WT observations: the background has a similar count rate as the source and the centroiding is difficult. Some slight systematic errors cannot be totally excluded but the consistency of the obtained data advocates for a limited impact on our results.

\begin{figure*}
  \begin{center}
    \includegraphics[width=8cm]{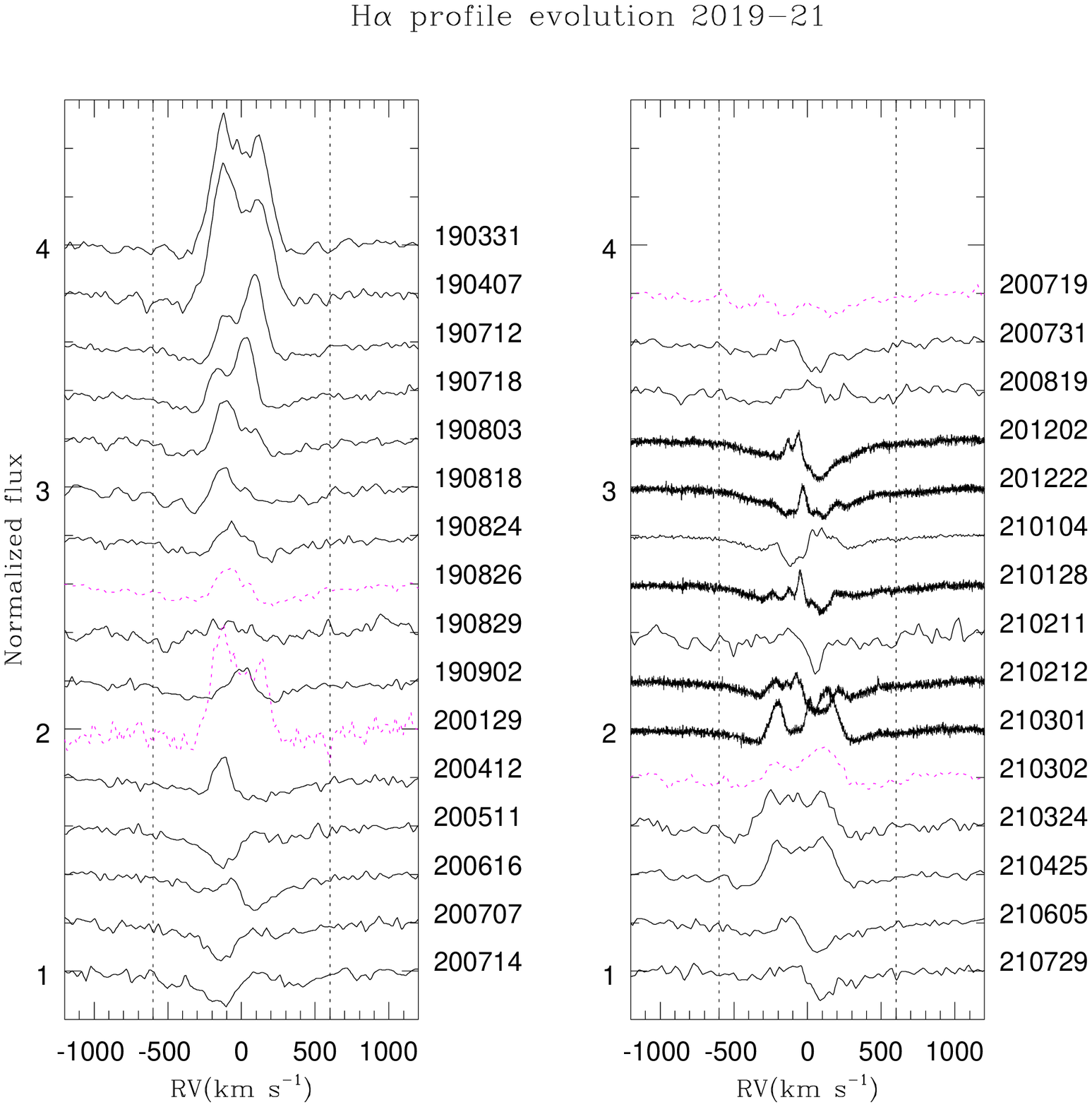}
    \includegraphics[width=8cm]{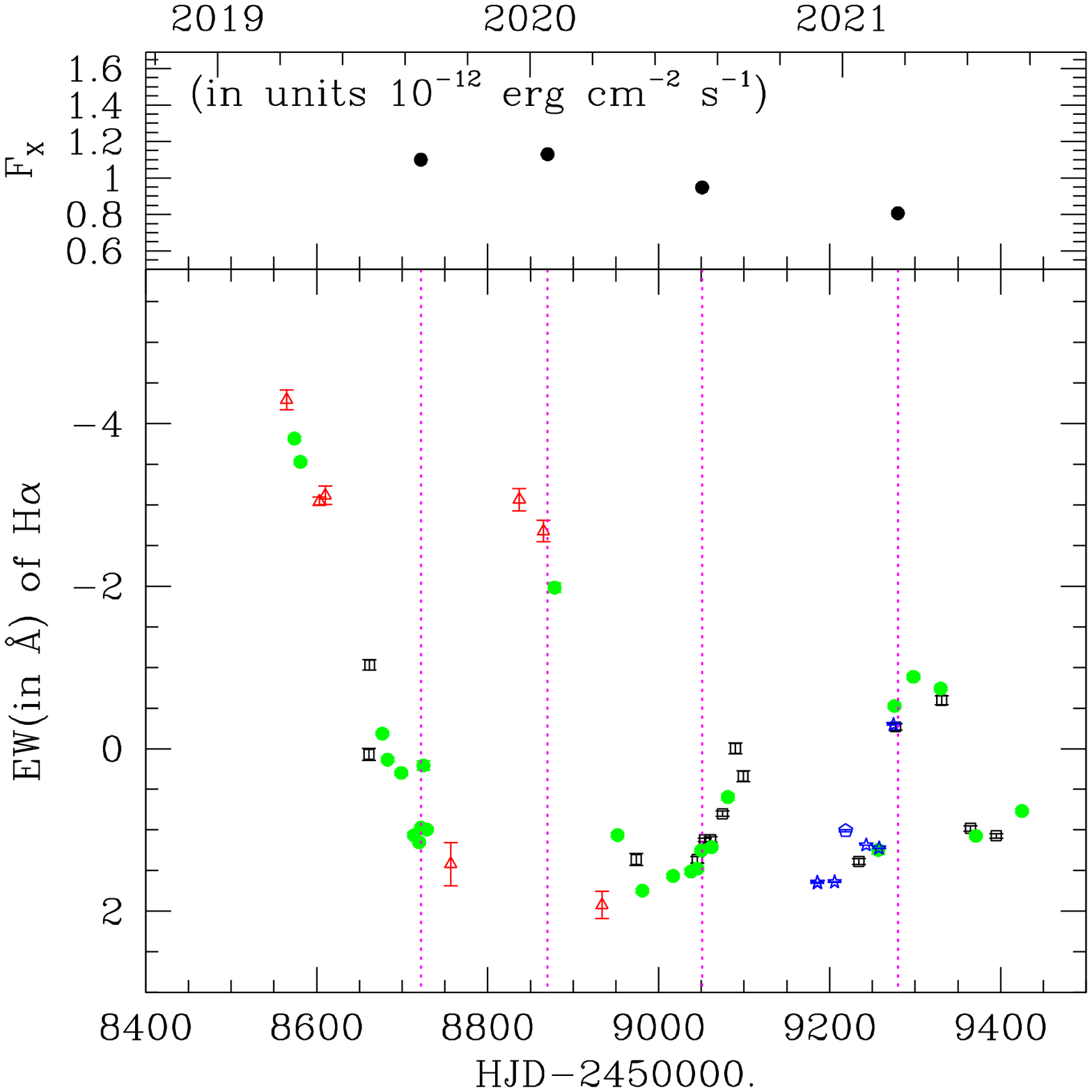}
  \end{center}
  \caption{{\it Left panels:} Evolution with time of the profile of the H$\alpha$ line observed in \hd\ during our monitoring campaign. Dates in YYMMDD are provided to the right of the line. Spectra taken close to the time of an \xmm\ observation are shown with a magenta dotted line, while vertical black dotted lines indicate the interval chosen for equivalent width determination. Note that the noisier ($SNR<20$) or lower-resolution($R<3000$) spectra are not shown. {\it Right panel:} Evolution with time of the H$\alpha$ equivalent widths measured in the --600 to +600\,\kms\ interval and of X-ray fluxes. The top axis provides the date in years while the bottom axis uses julian dates. Red open triangles display the values obtained from the noisier spectra ($SNR<20$), black open squares those measured on the lower resolution data ($R<3000$ but $SNR>20$), green dots those derived from other amateur spectra ($R>3000$ and $SNR>20$), and blue symbols those measured on ESO spectra (stars for UVES and open pentagon for X-Shooter). Note the very good agreement between ESO and amateur data taken at similar dates. The vertical magenta dotted lines indicate the times of the \xmm\ observations. }
\label{profhd}
\end{figure*}

\begin{figure}
  \begin{center}
    \includegraphics[width=8cm]{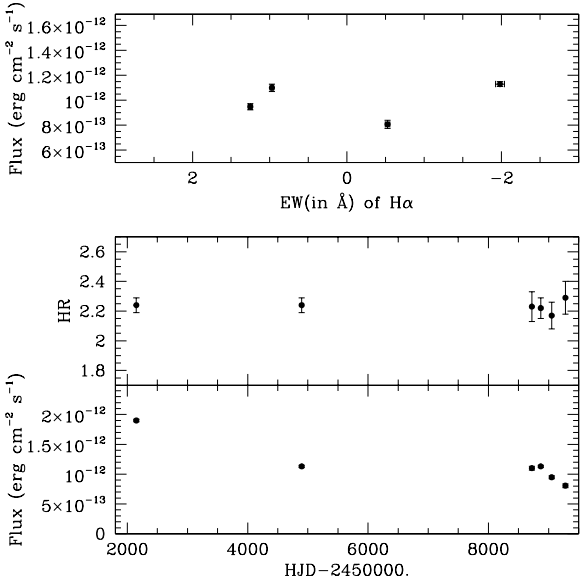}
  \end{center}
  \caption{X-ray fluxes and hardness ratios of \hd\ over time (bottom) and compared to $EW$ (top). }
\label{xparhd}
\end{figure}

Spectra were then fitted in {\sc Xspec} v12.11.1 using absorbed optically thin thermal emission models with solar abundances from \citet{asp09}. For \xmm, all EPIC spectra (pn, MOS1, MOS2) were fitted simultaneously; for {\it Chandra}, zeroth order spectra were fitted for individual exposures while a fit to the HEG and MEG grating spectra combining both orders and all exposures was also made.  The chosen models were as in \citet{naz18}. Results are provided in Table \ref{fitsx} - slight differences with those reported in \citet{naz18} for the same exposures come from the improved atomic parameters of the fitting tool. For \vc, a single temperature fit could a priori provide sufficient results for {\it Swift} spectra. Nevertheless, two-temperature fits were also tried to ease the comparison with \xmm\ results. As the {\it Swift} data have lower quality, especially at low energies, the temperatures were fixed to those found with \xmm, which show little changes between the two observations. However, even with fixed temperatures, the strength of the low temperature component could not be constrained. We therefore further fixed the ratio between the normalization factors of the two thermal components to the value observed with \xmm\ in 2021.  

\begin{table*}
  \caption{Best-fit models to the X-ray spectra. \label{fitsx}}
  \begin{tabular}{lcccccccccc}
    \hline
ID   & I & $HJD$ & $N_{\rm H}^{ISM}$ & $N_{\rm H}$ & $kT$ & $norm$ & $\chi^2$/dof & $F_{\rm X}^{obs}$ & $F_{\rm X}^{ISM-cor}$ & $HR$ \\
     &   & $-2450000.$ &\multicolumn{2}{c}{($10^{22}$\,cm$^{-2}$)} & (keV) & ($10^{-4}$\,cm$^{-5}$) & & \multicolumn{2}{c}{(tot, $10^{-12}$\,erg\,cm$^{-2}$\,s$^{-1}$)} & \\
\hline                                               
\multicolumn{11}{l}{\hd}\\
0315  &x&2149.849 &0.2 &0.020$\pm$0.008& 7.8$\pm$0.4 &11.8$\pm$0.13 &518.71/468 &1.90$\pm$0.02 &2.09 &2.24$\pm$0.05\\
8929  &c&4817.692 &0.2 &0.00$\pm$0.02  &10.3$\pm$2.2 &11.3$\pm$0.39 &49.52/56   &1.90$\pm$0.12 &2.10 &2.44$\pm$0.27\\
10835 &c&4820.076 &0.2 &0.00 (fixed)   & 9.4$\pm$1.9 &8.45$\pm$0.33 &49.68/45   &1.41$\pm$0.08 &1.55 &2.35$\pm$0.22\\
10834 &c&4821.515 &0.2 &0.00$\pm$0.03  &12.5$\pm$2.7 &8.58$\pm$0.24 &106.86/89  &1.47$\pm$0.07 &1.60 &2.61$\pm$0.19\\
10836 &c&4822.342 &0.2 &0.06$\pm$0.07  &12.9$\pm$6.0 &10.2$\pm$0.60 &39.87/51   &1.70$\pm$0.12 &1.84 &2.91$\pm$0.33\\
comb. &c&4820.201 &0.2 &0.13$\pm$0.13  &12.5$\pm$8.2 &12.0$\pm$0.66 &128.98/444 &1.97$\pm$0.12 &2.12 &3.21$\pm$0.38\\
1692  &x&4897.283 &0.2 &0.009$\pm$0.008& 8.1$\pm$0.5 &6.97$\pm$0.09 &467.69/448 &1.13$\pm$0.02 &1.25 &2.24$\pm$0.05\\
3610  &x&8721.890 &0.2 &0.000$\pm$0.007& 8.3$\pm$0.8 &6.74$\pm$0.11 &283.72/268 &1.10$\pm$0.03 &1.22 &2.23$\pm$0.10\\
3684  &x&8870.137 &0.2 &0.000$\pm$0.005& 8.3$\pm$0.5 &6.94$\pm$0.09 &396.86/363 &1.13$\pm$0.02 &1.26 &2.22$\pm$0.07\\
3775  &x&9050.958 &0.2 &0.00 (fixed)   & 7.8$\pm$0.7 &5.86$\pm$0.10 &270.94/265 &0.95$\pm$0.03 &1.05 &2.17$\pm$0.09\\
3890  &x&9279.849 &0.2 &0.000$\pm$0.008& 8.8$\pm$1.1 &4.88$\pm$0.10 &265.48/234 &0.81$\pm$0.03 &0.89 &2.29$\pm$0.11\\
\hline                                               
\multicolumn{11}{l}{\vc}\\
1306  &x&4126.327 &0.043&0.065$\pm$0.013& 0.27$\pm$0.03 &1.26$\pm$0.37 & 494.27/443 & 3.09$\pm$0.05 & 3.18 & 1.97$\pm$0.06\\
      &         &     &               & 6.48$\pm$0.33 &18.5$\pm$0.28 \\
3967  &x&9434.160 &0.043&0.080$\pm$0.012& 0.28$\pm$0.03 &1.60$\pm$0.39 & 560.49/451 & 3.28$\pm$0.05 & 3.38 & 1.89$\pm$0.06\\
      &         &     &               & 6.02$\pm$0.37 &20.0$\pm$0.30 \\ 
00014422001  &s&9409.837 &0.043&0.21$\pm$0.09& 0.28\&6.02 (fixed)& 25.3$\pm$2.5 &56.46/51 & 3.85$\pm$0.32 &3.93 & 2.37$\pm$0.32\\
00014422002  &s&9428.756 &0.043&0.10$\pm$0.10& 0.28\&6.02 (fixed)& 24.5$\pm$3.3 &30.19/34 & 3.97$\pm$0.41 &4.08 & 1.96$\pm$0.35\\
00014422003+4&s&9446.647 &0.043&0.00$\pm$0.10& 0.28\&6.02 (fixed)& 16.6$\pm$2.4 &37.16/36 & 2.88$\pm$0.43 &2.98 & 1.61$\pm$0.39\\
00014422005  &s&9465.858 &0.043&0.09$\pm$0.10& 0.28\&6.02 (fixed)& 23.7$\pm$3.0 &36.50/38 & 3.88$\pm$0.36 &3.99 & 1.91$\pm$0.31\\
00014422006  &s&9486.837 &0.043&0.00$\pm$0.08& 0.28\&6.02 (fixed)& 20.9$\pm$2.6 &50.01/48 & 3.62$\pm$0.40 &3.76 & 1.62$\pm$0.34\\
00014422007  &s&9561.986 &0.043&0.00$\pm$0.11& 0.28\&6.02 (fixed)& 14.9$\pm$2.4 &51.76/39 & 2.59$\pm$0.36 &2.68 & 1.63$\pm$0.33\\
00014422008  &s&9583.002 &0.043&0.00$\pm$0.07& 0.28\&6.02 (fixed)& 16.6$\pm$2.1 &35.38/32 & 2.88$\pm$0.38 &2.99 & 1.62$\pm$0.35\\

\hline                                               
\multicolumn{11}{l}{$\pi$\,Aqr}\\
00010659001-39&s&8344.345 &0.036&0.53$\pm$0.06 &27.6$\pm$9.4 &105.5$\pm$4.6 &396.03/436 &16.0$\pm$0.4 &16.1 & 6.10$\pm$0.24\\
\hline
  \end{tabular}
  
{\scriptsize Fitted models were of the form tbabs$\times$phabs$\times$apec, with the first absorption fixed to the interstellar value. ID refers to the revolution number (for \xmm) or ObsID (for {\it Chandra}; ``comb'' indicates the fitting of the fully combined (all obs, both orders) {\it Chandra} HEG and MEG grating spectra. Column I identifies the facility used (x for \xmm, c for {\it Chandra}, s for {\it Swift}). The hardness ratios are defined by $HR = F_{\rm X}^{ISM-cor}(hard)/F_{\rm X}^{ISM-cor}(soft)$, with $F_{\rm X}^{ISM-cor}$ the flux after correction for interstellar absorption and soft and hard energy bands being defined as 0.5--2.0 keV and 2.0--10.0 keV, respectively (the total band being 0.5--10.0 keV). Errors correspond to 1$\sigma$ uncertainties; they correspond to the larger value if the error bar is asymmetric. For {\it Swift} spectra of \vc, the normalization factor of the 0.28\,keV component is fixed to 0.08 times that of the 6.02\,keV component, as in \xmm\ data of Rev. 3967, and only the normalization of that hottest component is provided here.}
\end{table*}

\section{Results}

\begin{figure*}
  \begin{center}
    \includegraphics[width=8.8cm]{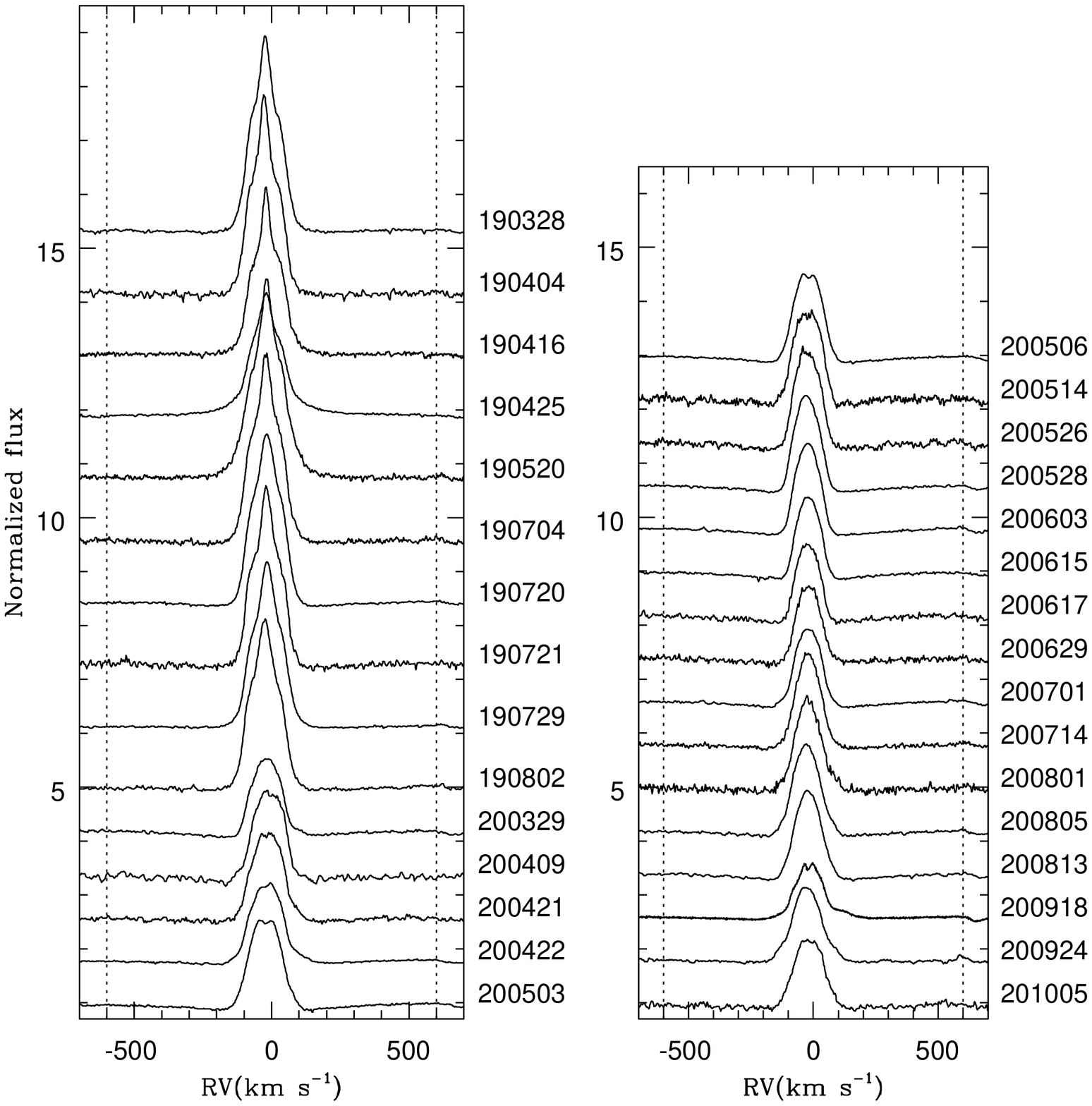}
    \includegraphics[width=8.8cm]{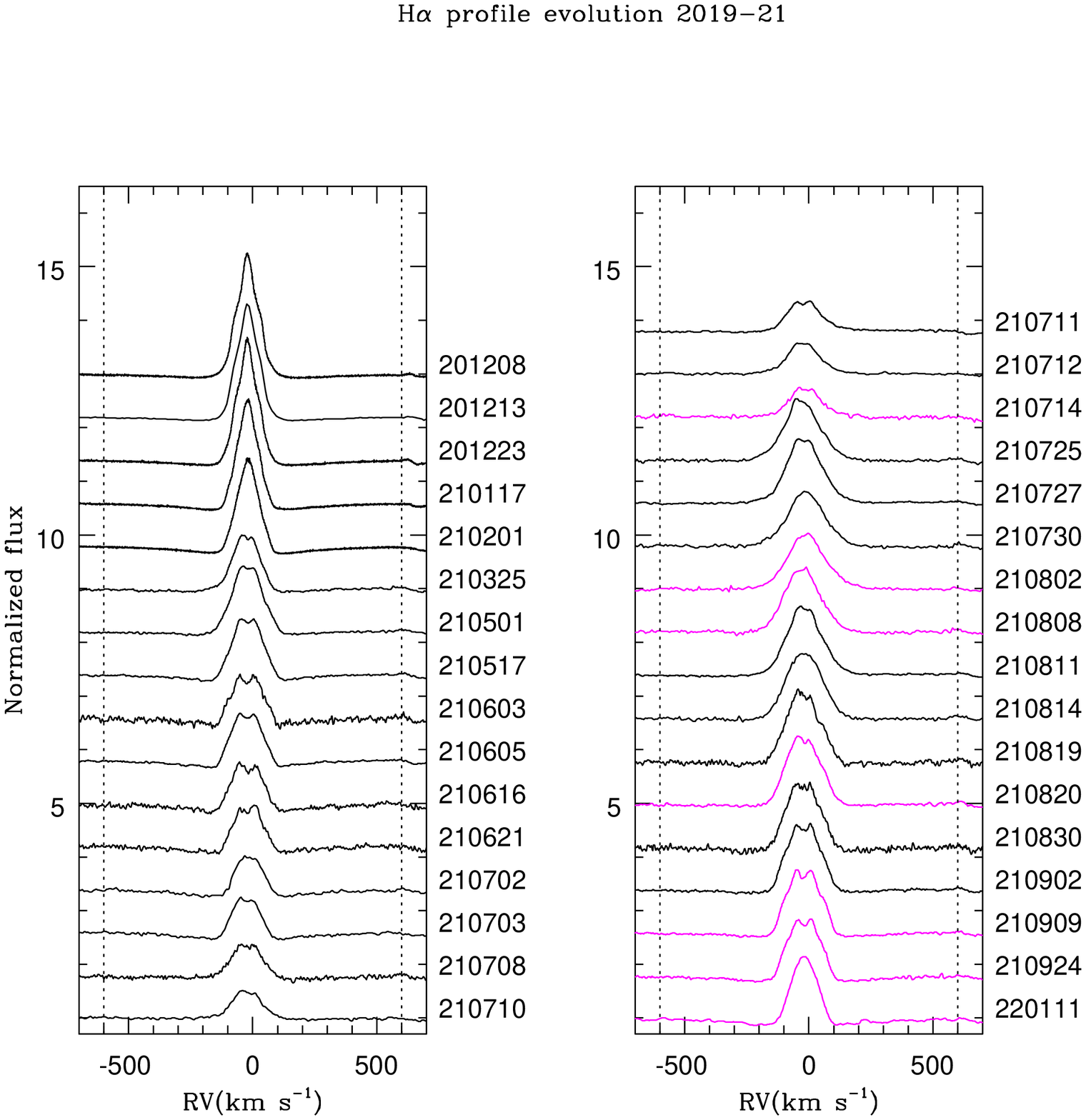}
  \end{center}
  \caption{Same as left panel of Fig. \ref{profhd} but for the profiles of \vc. Note that low resolution spectra here have $R<5000$. }
\label{profvc}
\end{figure*}

\begin{figure}
  \begin{center}
    \includegraphics[width=8cm]{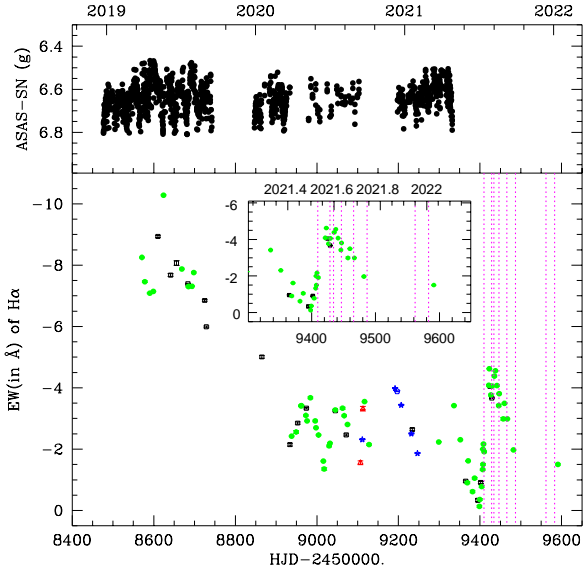}
  \end{center}
  \caption{{\it Top:} g-magnitudes of \vc\ recorded by ASAS-SN. {\it Bottom:} Same as the right panel of Fig. \ref{profhd} but for \vc. Note that low resolution spectra are here defined as $R<5000$. }
\label{profvc2}
\end{figure}

\begin{figure}
  \begin{center}
    \includegraphics[width=8cm]{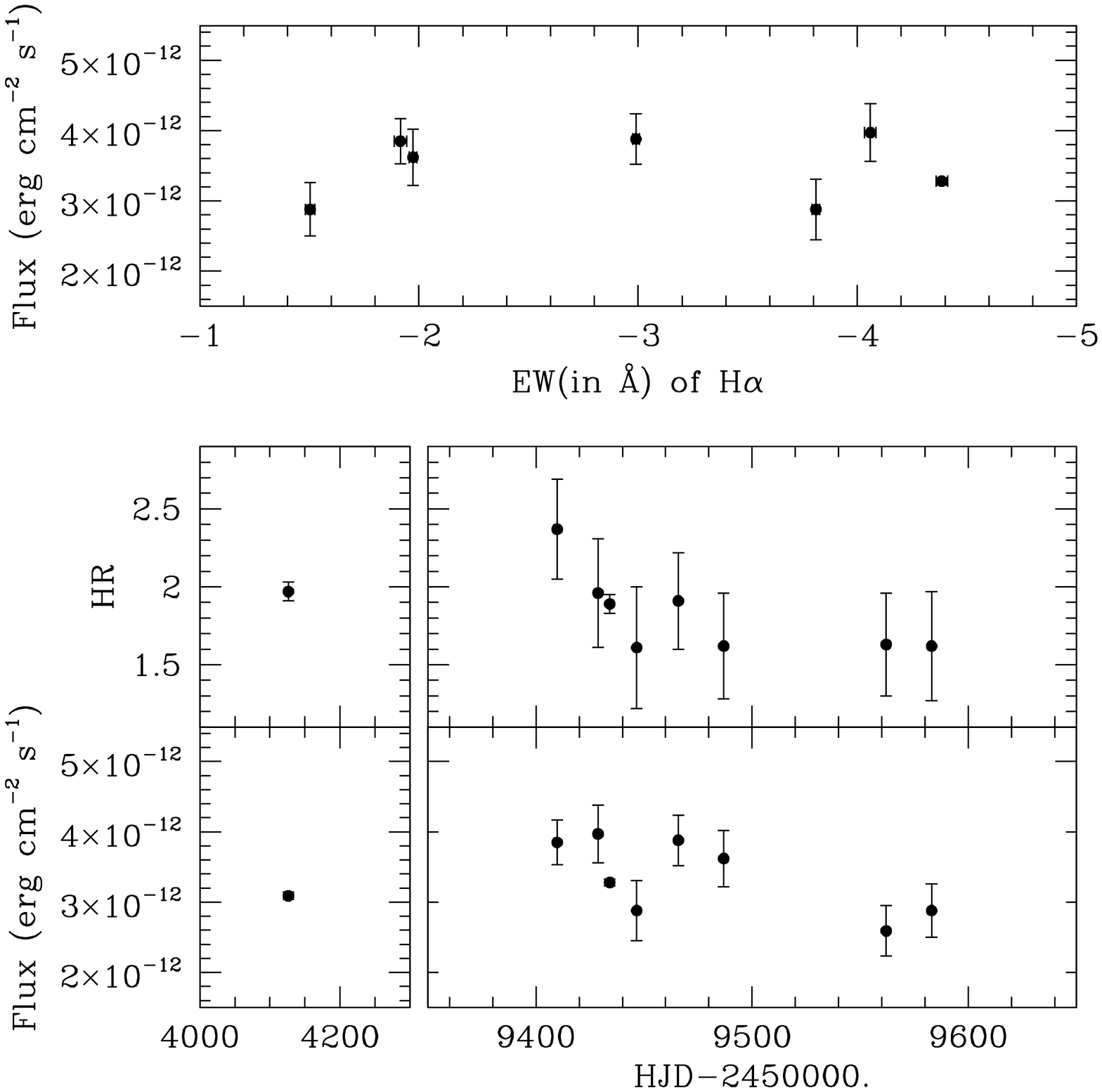}
  \end{center}
  \caption{Same as Fig. \ref{xparhd} but for \vc. }
\label{xparvc}
\end{figure}

\subsection{\hd}
When the optical monitoring began in 2019, the H$\alpha$ line of \hd\ displayed a double-peaked emission, with no trace of absorption. The equivalent width was moderate (--4.3\,\AA, a slightly lower value than in an older - May 2017 - spectrum). Over the course of the year, the emission steadily decreased, with the underlying absorption more and more clearly detectable. A moderate emission suddenly re-appeared in January 2020 ($EW\sim-3$\,\AA), but disappeared soon afterwards. The H$\alpha$ line then remained in absorption, although a small emission did shortly re-appear in December 2020 and March-April 2021 (Fig. \ref{profhd}). Thus, during the monitoring, the disk did not undergo a monotonic disappearance. Rather, the behaviour appears somewhat erratic, with an overall disappearance trend superimposed on temporary disk reinforcements, probably corresponding to small mass ejection events. The complex emission line profile, with changing width and even multiple peaks as revealed by ESO spectra, also points towards a complicated disk geometry at those times: the ejected material does not seem to form a single blob slowly and smoothly mixing with the disk.

The H$\alpha$ line profile was very different when the various \xmm\ observations were taken (see magenta profiles in Fig. \ref{profhd}). In August 2019, the absorption was dominant and the emission weak. In January 2020, the emission was moderate while, in July 2020, there was no trace of emission. Finally, in March 2021, the emission was intermediate between the first two cases. There is unfortunately no information on the shape of the H$\alpha$ line profile at the time of the older \xmm\ or {\it Chandra} observations.

During each X-ray observation, short-term flux variations can be spotted in the lightcurves, as are common in \gc\ stars \citep[e.g.][]{smi12}. It should however be noted that the count rate does not allow to make lightcurves with extremely short time bins (e.g. 1\,s) thought to be typical of the ``shot''/flaring emission of \gc\ stars hence such timescales remain unexplored for \hd. The Appendix shows these \xmm\ lightcurves. It reveals that the hardness of the emission, estimated from the ratio of hard (2--10\,keV) to soft (0.5--2\,keV) count rates, does not seem to change in a significant way over the whole \xmm\ dataset. With a Pearson correlation coefficient of only 0.33, the hardness also does not appear significantly correlated to the strength of the X-ray emission, estimated from the full band (0.5--10\,keV) count rate. Finally, it may be noted that the dispersion of the lightcurve points remains similar if one considers a single exposure or the whole dataset, despite different values of the average count rate.

Table \ref{fitsx} provides the results of spectral fits. We can see that there is some systematic difference between \xmm\ and {\it Chandra} results. This is in part due to the stellar variations, but also to remaining cross-calibration problems: even for a stable object, it is quite common to not get exactly the same modelling results. Note that the comparison between the combined grating data and the individual 0th order spectra shows a good agreement, indicating little effects of pile-up which could potentially still affect the latter data.

The X-ray flux was about $2\times10^{-12}$\,erg\,cm$^{-2}$\,s$^{-1}$ at the time of the oldest observations, but it then decreased by a factor of two in more recent years. Focusing on our four monitoring observations, the star did not appear at its lowest flux when the H$\alpha$ emission was the lowest (absorption only, July 2020) nor did it appear brighter in January 2020 or March 2021 when a small surge in disk emission occurred. Rather, there seems to be a slow, monotonic flux decline ($\sim$30\%) over the entire monitoring interval. Furthermore, restricting to comparing \xmm\ spectra to avoid any cross-calibration problems, we see that the ratio $HR$ between hard and soft fluxes (after correction for interstellar absorption) remained stable at all times, while the temperature and absorption also agree within errors. Clearly, the \gc\ characteristics did not disappear, nor did they change of relative strength, even though the flux changed and the disk evolved. Figure \ref{xparhd} graphically displays this evolution of flux and hardness ratios over time. The top panel of the same figure compares the fluxes to the H$\alpha$ line strength: it confirms the lack of correlation between X-ray parameters and the H$\alpha$ line.
 
\subsection{\vc}
At the start of the monitoring in March 2019, the star displayed a rather strong emission, with the line amplitude nearly reaching four times the continuum level (Figs. \ref{profvc} and \ref{profvc2}, $EW\sim-8$\,\AA). The line profile had a single peak, although with prominent shoulders (i.e. the line had a winebottle shape). The emission slowly decreased in April, then went shortly back to its initial level in May, before resuming its decrease. After this downward trend, the emission appeared to stabilize the next year. Only small variations were seen around an average equivalent width of $\sim$--3\,\AA. At that time, the line profile reached an amplitude of only 1--1.5 times the continuum level. Furthermore, the shoulders seemed to have disappeared while the broad stellar absorption started to appear in the high-velocity wings of the profile. The high-resolution ESO spectrum taken in September 2020 shows that the single peak seen in amateur data is actually made of two very close subpeaks. In December 2020, the high-resolution ESO data revealed that the emission became narrower but of larger amplitude ($EW=-4$\,\AA, amplitude twice the continuum level, with clear hints of the photospheric absorption outside the emission range). The emission gradually decreased over the next months, down to $EW=-2$\,\AA, showing that the disk re-building was only temporary. In parallel, the separation between peaks in the double-peaked H$\beta$ line profile changed from 18.5\,\kms\ in early December 2020 to 52\,\kms\ at the beginning of February 2021. Using Eq. 2 of \citet{zam19}, this translates into a size change for the emission region by a factor of $\sim$8 in just two months. Such a variation in peak separation is typical of Be disks as a larger disk reaches lower orbital velocities at its periphery than a small disk closer to the star \citep{hum95,zam19}.

The emission then stabilized in March--May 2021, to finally resume its slow decrease in June. The absorption wings were clearly visible from mid-May to early July. While one would have expected the emission to slowly disappear, the absorption wings suddenly filled up. This suggests an input of fresh material close to the star, where velocities are large. In such a case, it is expected that the material will gradually spread out over the whole disk: indeed, the core of the emission soon became broader and stronger, reaching one time the continuum level at the end of July (compared to an amplitude of one-half in mid-July). After this sudden event, the emission traced by the H$\alpha$ line resumed its decrease. 

Contrary to \hd, the H$\alpha$ line was never largely dominated by absorption in our observations of \vc, a clear emission being visible in all spectra, even at lowest $EW$. Both stars however display irregular decreases, often interrupted by short flaring episodes. Since the July 2021 flaring led to an equivalent width increase never seen before in our monitoring, we decided to trigger X-ray observations to follow the reaction to the reinforcement of the disk, rather than to its disappearance as in \hd. {\it Swift} was the fastest to react, with the first exposure taken less than a week after the event, while \xmm\ data were taken a month later. 

Comparing first the highest quality data, it is obvious that the 2007 and 2021 \xmm\ spectra display very similar properties (absorption, temperatures, overall luminosity, hardness ratio, see Table \ref{fitsx}). Although the disk state in 2007 is not known, it would be quite a coincidence for it to be the exact same one as in 2021. In addition, each \xmm\ lightcurve displays short-term flux variations, as found for \hd\ and other \gc\ stars, but no hardness variations (see Appendix). Again, there seems to be no correlation between the overall count rate and the hardness in these lightcurves. Finally, while the disk emission was much lower in 2021 than at the beginning of the optical monitoring in 2019, the \gc\ character remained clear. 

The {\it Swift} data allow to examine the behaviour of the star in 2021 over a longer timescale, albeit with lower quality data. \vc\ appears slightly brighter in the second exposure and slightly fainter in the last ones, but the changes remain within 2$\sigma$: again, the star seems to display a rather similar X-ray emission at all times. These results are shown in Fig. \ref{xparvc}, which also graphically demonstrates the absence of correlation between X-ray flux and strength of the H$\alpha$ line.

Finally, {\it ASAS-SN} provided photometry over a long timescale\footnote{\vc\ was also observed by {\it TESS} during our monitoring campaign and these observations are reported in \citet{naz20tess}. Significant variations of about 0.2\,mag were recorded by {\it TESS} over its few weeks' observing window and they are confirmed by {\it ASAS-SN} data.} (see top panel of Fig. \ref{profvc2}). Despite their noise, those data clearly show that the broad-band photometry remains rather stable while the H$\alpha$ line strongly changes.

\section{Discussion}
\subsection{Observed changes in \gc\ analogs}
Amongst \gc\ analogs, only three stars had previously been followed through a transition of their disk. The Oe star HD\,45314, the hottest \gc\ analog known so far, was monitored extensively in the optical range \citep{rau18}. X-ray observations were obtained as the star displayed very different disk states, as traced by the H$\alpha$ line: strong emission (equivalent width close to --23\AA), shell phase, and very small emission (equivalent width of --7.9\AA). Changes in $V$-band photometry were also detected, indicating that even the inner parts of the disk were affected by the variation. Between the first and last observations, the X-ray flux was reduced by an order of magnitude and the hardness of the spectrum markedly decreased too ($HR$ changed from 4.3 to 1.8, see \citealt{naz18}). The \gc\ character was thus disappearing, suggesting a direct link between the disk and the generation of X-rays \citep{rau18}. 

In contrast, the monitoring of $\pi$\,Aqr drew a different picture \citep{naz19piaqr}. X-ray data were here also taken in two very different situations: as the disk had nearly completely disappeared (H$\alpha$ equivalent width of --1.7\AA) and as the emission associated to the disk was strong (equivalent width of --23\AA, with a disk size five times larger). Here too, simultaneous variations in broad-band photometry were recorded. The first X-ray observation was a single \xmm\ snapshot but the second observing campaign corresponds to a set of short exposures taken by {\it Swift} over 250\,d (i.e. three orbital cycles). There was no obvious relation between the X-ray parameters derived from individual {\it Swift} exposures and the orbital phase or between them and the disk fluctuations of $\pi$\,Aqr as traced by H$\alpha$ line strength \citep{naz19piaqr}. To readdress this issue, we now combine the {\it Swift} exposures with the on-line tool\footnote{https://www.swift.ac.uk/user\_objects/} to get a single, higher-quality spectrum. We fit it in the same way as done for the \xmm\ spectrum in \citet{naz18}, see Table \ref{fitsx} for results. While the short {\it Swift} exposures show individual fluxes varying by a factor of a few, as usual for the short-term intrinsic variations of \gc\ stars \citep{naz19piaqr}, the flux of the combined {\it Swift} dataset appears to be $\sim$50\% higher than in the \xmm\ data. This change is due to the increase of the hard X-ray flux, leading to a variation of the hardness ratio $HR$ from 3.6 to 6.1. However, it is important to stress that the hard component was very clear and strong at all times. The \gc\ character thus never disappeared for $\pi$\,Aqr.

The third \gc\ star that has been simultaneously monitored at optical and X-ray wavelengths is \gc\ itself. \citet{mot15} reported a good correlation, without any time lag, between averages over optical observing seasons of X-ray fluxes measured by {\it RXTE} and of disk brightnesses as traced by $V$-band magnitudes. Moreover, variations with timescales near 70\,d were found in both wavelength ranges. Recently, Rauw et al. (in prep) reported on a set of \xmm\ observations taken during an eruption event (with H$\alpha$ equivalent width transitioning from --45\AA\ to nearly --55\AA\ and then back to the initial value). They compared the high-energy emission then observed to that recorded in older \xmm\ observations when the disk emission was significantly lower (H$\alpha$ equivalent width ranging from --27 to --35\AA). Both flux and hardness variations are observed in the X-ray range. Besides the usual short-term ``flaring'' activity, the mean flux of each \xmm\ observation varied by a factor of two. The hardness ratio was similar in six out of the ten observations ($HR\sim 3$) and three more observations displayed values rather close to that, despite the different strength of the H$\alpha$ line at these nine dates. However, the X-ray spectrum clearly was hardest ($HR\sim 8$) and faintest at the time of the maximum emission. This variation stems from the soft X-rays (the soft flux changed by a factor of 3--4), most probably because of a larger absorption. Such characteristics are reminiscent of the localized absorbing events reported by \citet{ham16} and \citet{smi19}, of which this event would then be an extreme case. Indeed, the next exposure, taken only a month later when the H$\alpha$ emission was still very strong, did not display those features. It can nevertheless not be totally excluded that the observed change is a direct high-energy reaction to the disk event. In any case, there was no obvious correlation between the H$\alpha$ emission and the X-ray properties during this event or at previous epochs of \xmm\ observations. Finally, it may be noted that no significant change in $V$-band photometry was detected during this recent emission event.

\hd\ and \vc\ add to this picture of constrasting behaviours. The X-ray flux of \hd\ displayed long-term variations by a factor of two, but the hardness of the emission remains stable ($HR\sim2.2$ in all \xmm\ data). In particular, the \gc\ characteristics were still present when the optical spectrum showed no trace of emission (H$\alpha$ equivalent width of 1.3\AA). For \vc, the disk never disappeared entirely: when it seemed on the edge of doing so (at an H$\alpha$ equivalent width of $\sim-0.2$\,\AA), the H$\alpha$ emission suddenly increased. At X-ray wavelengths, however, the spectral properties show little change, both in flux and hardness. 

In summary, three disk disappearances (or near-disappearances) were monitored, for HD\,45314 (O9pe), $\pi$\,Aqr (B1Ve), and \hd\ (B0Ve). In all cases, a lower X-ray flux than observed before was measured at these times. However, the amplitude of the flux change wildly varied (an order of magnitude for HD\,45314, about a factor of two for the other two). The hardness variations are even more different: HD\,45314 had such a low hard X-ray emission that it formally lost its \gc\ character, the hardness of $\pi$\,Aqr decreased but the star still fulfilled the criteria to remain a \gc\ analog, and \hd\ kept the hardness measured before. This suggests the existence of a link, but a loose one, between the H$\alpha$ emission of the disk and X-ray emissions.

In parallel, two disk ``flarings'' were monitored, for \gc\ and \vc. They were qualitatively very different, as the disk of \gc\ was already strong when it became stronger while that of \vc\ was on the edge of disappearance when a disk re-building was detected. However, apart from a single \xmm\ exposure of \gc, there was no large change of X-ray properties recorded in both stars. It is important to note that, for both stars, broad-band photometry showed no specific trace of flaring at the time of the recorded events (see above for \vc, and see Rauw et al. in prep for \gc). 

\subsection{The case of Be-XRBs}
What do these monitoring results tell us on the generation of X-rays in \gc\ stars? To shed light on this issue, a comparison with the usual behaviour of Be stars in X-ray binaries (XRBs) should be done. In such systems, the X-ray variability is often classified in two categories. Type I outbursts occur at specific orbital phases, when accretion is enhanced at periastron passages, while type II outbursts are rather linked to major disk changes, often after some reaction delay \citep[e.g.][]{gru07,cam12,lut12,alf17}. Most systems remain X-ray quiet when the Be disk is small or has disappeared \citep{neg01} although a few systems have undergone type II outbursts even in such conditions \citep{mon17}.

The X-ray observations of \gc\ analogs have not revealed any outburst up to now, hence their closer analogs amongst X-ray binaries may be the low-eccentricity systems such as X\,Per. From a long monitoring of this system, \citet{zam19} found a direct correlation, without time lag, between the X-ray flux and the equivalent width of the H$\alpha$ line. However, this occurred only when the emission was very strong, i.e. for a large disk. At lower emission levels, the situation was clearly different, with the $V$-band photometry varying first, then the $EW$(H$\alpha$), and at last the X-ray flux, after years of delay. This can be understood recalling that the accretion material comes from the outer parts of the disk: if the disk is small, it will take some time for such an inner variation to change the accreting conditions near the companion while any variation at the periphery of a large disk, hence closer to the companion, will have a direct impact on the accretion flow. 

\subsection{Constraints on the \gc\ phenomenon}
The \gc\ monitorings provide two important clues. First, the \gc\ character remained for $\pi$\,Aqr, \hd, and \vc\ even when the H$\alpha$ emission was very weak. Of course, the current H$\alpha$ data does not enable us to conclude that the disk had entirely disappeared, even in the case of \hd. Indeed, though H$\alpha$ is certainly a powerful diagnostics for the presence of a circumstellar disk, the line forms over a rather wide radial extent and is much less sensitive to the innermost and outermost parts of the disk. However, it is certain that the disk size was much reduced at these times. In fact, the size of the disk $R_d$ can be evaluated from the peak separation $\Delta(V)$ in double-peaked profiles using $R_d=R_*\times (2\times v \sin(i)/\Delta(V))^2$ \citep{hua72,hum95,zam19}. The projected rotational velocities $v \sin(i)$ are 100 and 200\,\kms\ for \vc\ and \hd, respectively \citep{zor16}. The peak separation were measured each time the profile appeared double-peaked and this yielded disk sizes of a few $R_*$ for \hd\ and 10--30\,$R_*$ for \vc. Comparable values were found for HD\,45314 \citep{rau18} and $\pi$\,Aqr \citep{naz19piaqr}. The disk sizes can also be evaluated from the $EW$s \citep[see Eq. (3) in][]{zam19} but the application of the formula can be debated for such small $EW$ values (the largest $|EW|$ measured here is around 10\AA). However, \citet{rei16} also found that the disk sizes must be small for such low $EW$. In our cases, the disks are thus smaller than usually found in active states of Be-XRBs \citep[see also Fig. 5 of][]{zam19}. In any case, any compact companion would need to be very close for significant accretion (hence significant hard X-ray emission) to occur: if $|EW|<10$\,\AA\ then $P_{orb}<50$\,d, see \citet{coe15}. For a Be star mass of 12\,M$_{\odot}$ and a Oe star mass of 20\,M$_{\odot}$, a neutron star mass of 2\,M$_{\odot}$, and a $P_{orb}<50$\,d, the velocity amplitudes would be larger than 15--20\,\kms, which is not detected for \hd, \vc, or HD\,45314 \citep{rau18,naz21}. For $\pi$\,Aqr, the known orbital period is 84\,d, with a disk much smaller than the orbital separation \citep[and references therein]{naz19piaqr}. The current data thus seem to disfavour scenarios involving accreting companions. 

Second, the X-ray emission seems to react differently depending on the extent of the disk changes. Indeed, optical photometry and spectroscopy probe different zones of the disk: the stellar photometry is more sensitive to the densest and innermost parts of the disk, while the H$\alpha$ line rather probes the disk over a larger region, up to its periphery. In \gc, HD\,45314 and $\pi$\,Aqr, variations of the X-ray emission were spotted when both broad-band photometry and H$\alpha$ line profile changed. In contrast, the recent ``events'' in \gc\ and \vc\ were detected through H$\alpha$ measurements but both X-ray properties and optical broad-band photometry remained unaffected. All this may be a hint that the hard X-ray emission that characterizes the \gc\ phenomenon is born in the inner disks of the Be stars, rather than at their periphery.

The observed optical and X-ray behaviours of \gc\ analogs therefore reveal that the \gc\ character may remain even if the disk size is much reduced and that changes in X-rays are usually seen only if broad-band photometric variations occurred. This is difficult to reconcile with a crucial role of distant companions, and seems to bring support to the magnetic star-disk interactions where X-rays are generated closer to the Be star \citep{rob02}. The differences observed between \gc\ analogs could then result from a range of reasons such as varying inclinations, stellar rotation, and stellar temperatures. For example, the hotter Oe star HD\,45314 should have a stronger wind and a faster disk disappearance \citep{kee16}. In contrast, the disk of the cooler Be stars would take more time to disperse, hence re-building events would have time to occur and a full disk disappearance may be more difficult to get. With the inner parts of the disk still in place, the \gc\ character would then still be observable. Nevertheless, no full modelling of the star-disk interaction is available yet, hence it is difficult to exactly quantify its adequacy. Future modelling should explore the specific impact of geometry and stellar properties. 

\section{Summary and conclusion}
The H$\alpha$ line in Be stars is considered as one of the main probes of their disks. We have monitored this line for two \gc\ analogs, \hd\ and \vc, for several years. Both stars displayed decreasing line strengths, although interrupted by several short re-building events. The discovery of this behaviour triggered X-ray observations, to assess the impact of the disk changes on the peculiar high-energy emission of those stars.

For \hd, the H$\alpha$ line was fully in absorption in mid-July 2020. In parallel, the X-ray flux slightly decreased between August 2019 and March 2021, with no change in hardness. The flux level was comparable to that recorded a decade before in a previous X-ray exposure. The \gc\ character remained clear at all epochs.

\vc\ was monitored during a disk re-building event and no significant change in X-ray flux or hardness was detected. In parallel, ASAS-SN photometry in the visible range also appeared to remain stable.

The limited reaction to large H$\alpha$ variations and the presence of the \gc\ character even with a weak H$\alpha$ line, coupled to a stable photometry, seems to disfavour scenarios involving an X-ray source located far out in the disk, close to a companion, and rather hints at an X-ray generation closer to the Be star.

These results brought important clues regarding the \gc\ phenomenon, but of course much remains to be done. In particular, the optical and X-ray monitorings should continue, to exclude large time delays and/or to analyse the X-ray behaviour during long (and complete) disk disappearances. Also, while the disk clearly plays some role in the appearance of the \gc\ phenomenon (since all \gc\ are Be stars), it remains to be clarified why most Be stars are {\it not} \gc\ in character. 

\section*{Acknowledgements}
We thank Dr N. Schartel for granting us a DDT observation of \vc, and the {\it Swift} team for their help. We also thank Myron Smith for his comments and our useful discussions. Y.N. and G.R. acknowledge support from the Fonds National de la Recherche Scientifique (Belgium), the European Space Agency (ESA) and the Belgian Federal Science Policy Office (BELSPO) in the framework of the PRODEX Programme (contracts linked to XMM-Newton and Gaia). ADS and CDS were used for preparing this document. 

\section*{Data availability}
The ESO, {\it Swift, Chandra}, and \xmm\ data used in this article are available in their respective public archives, while the Australian optical amateur spectra are available in the public BeSS database (http://basebe.obspm.fr/basebe/). The Brazilian amateur data are available upon reasonable request.

\appendix
\section{X-ray lightcurves}

\begin{figure*}
  \begin{center}
    \includegraphics[width=5.8cm]{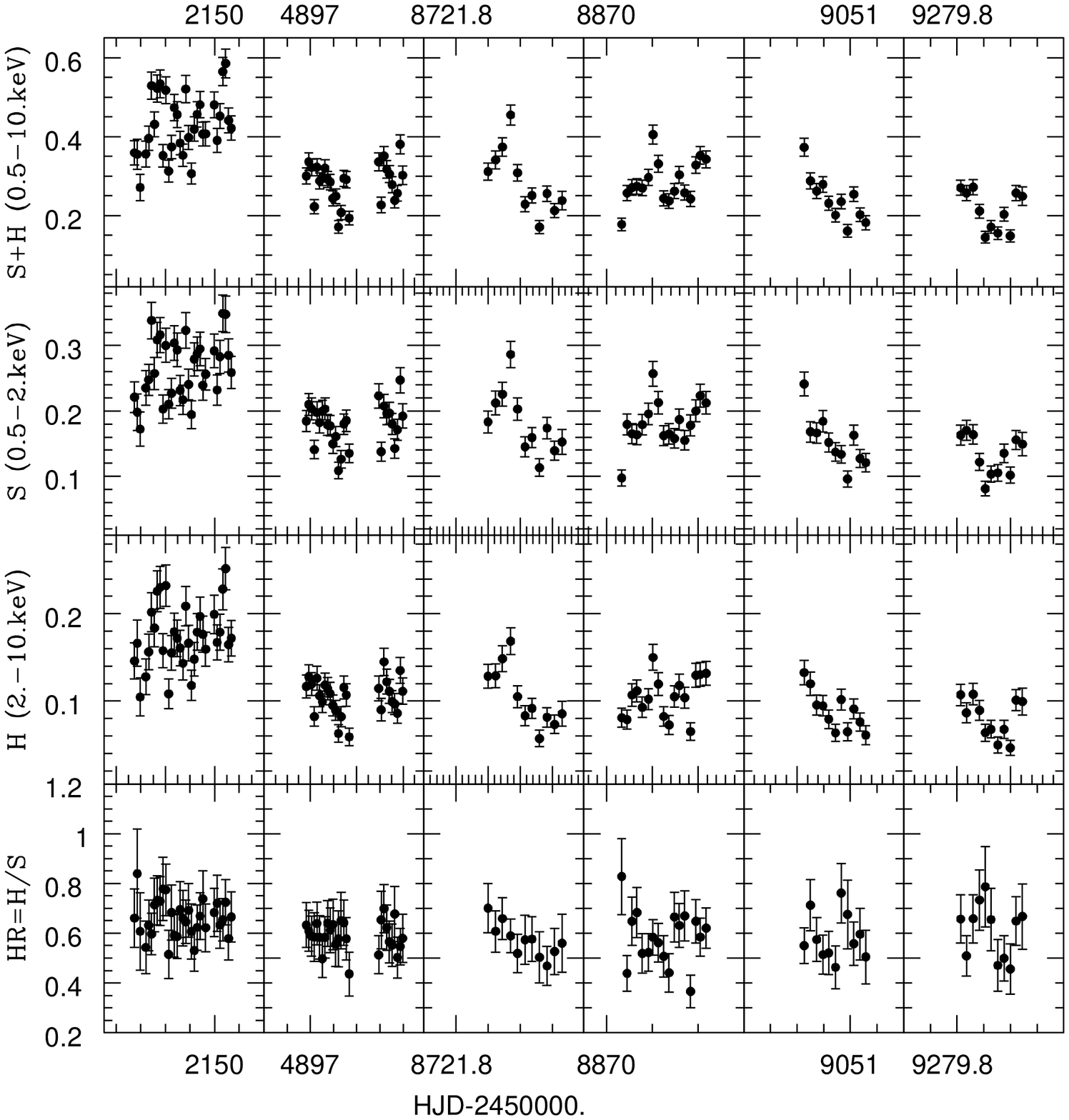}
    \includegraphics[width=5.8cm]{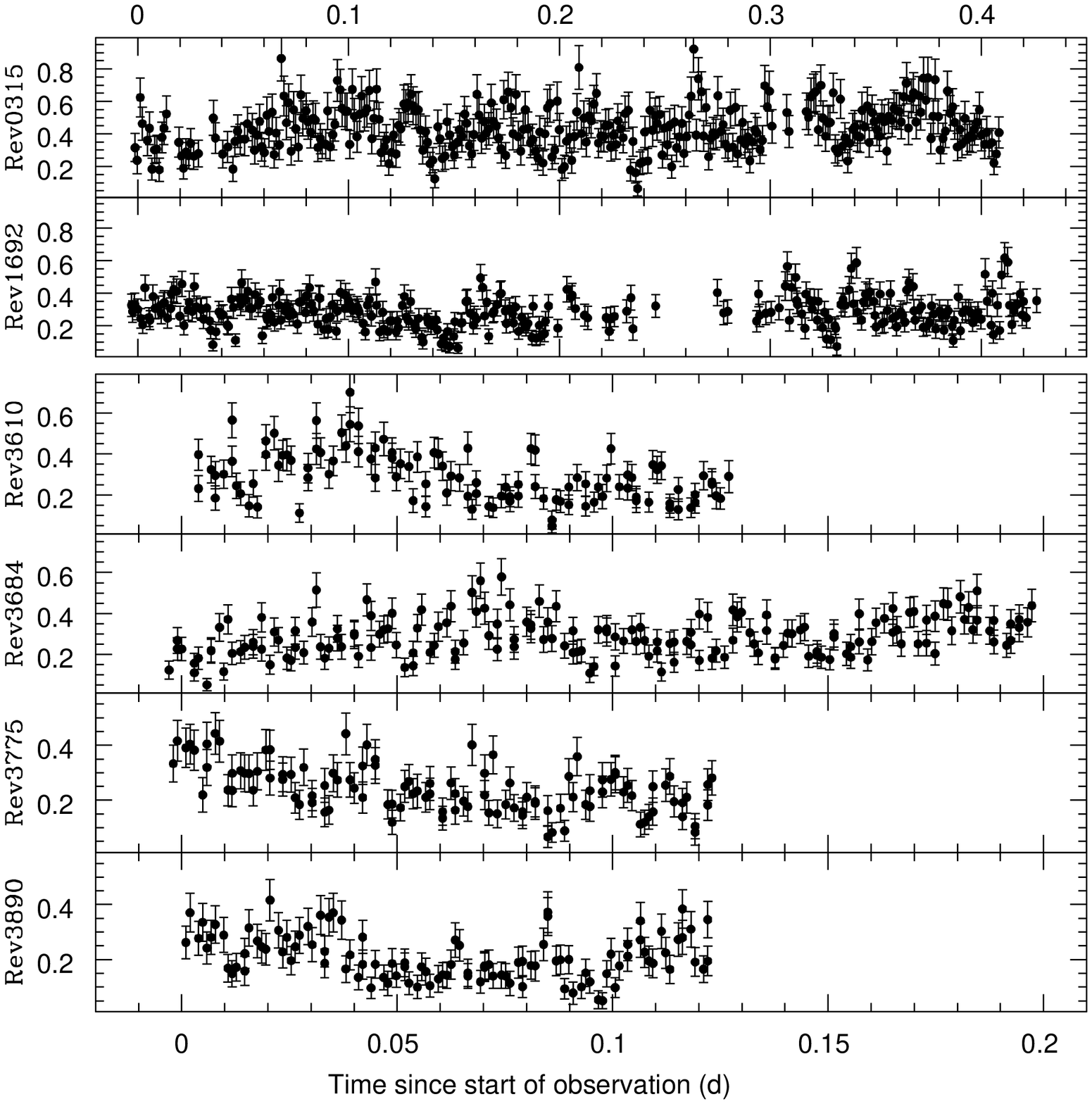}
    \includegraphics[width=5.8cm]{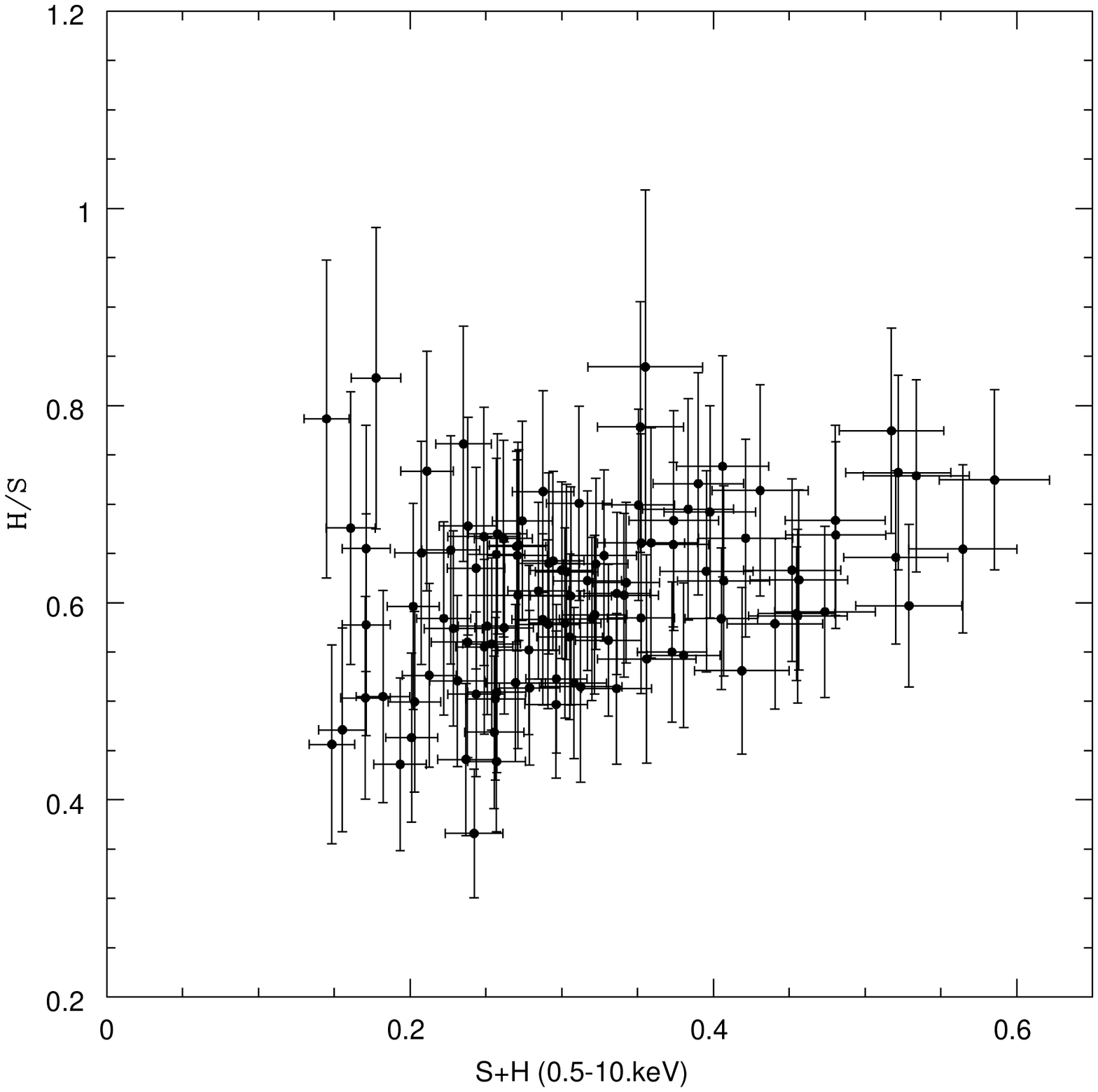}
  \end{center}
  \caption{{\it Left panels:} \xmm\ lightcurves of \hd\ in several energy bands and with a time bin of 1000\,s.  {\it Middle panels:} \xmm\ lightcurves of \hd\ in the total energy band and with a time bin of 100\,s. {\it Right panel:} Hardness ratio as a function of the total count rate in all \xmm\ lightcurves of \hd\ with a time bin of 1000\,s. }
\label{hdlcx}
\end{figure*}

\begin{figure*}
  \begin{center}
    \includegraphics[width=5.8cm]{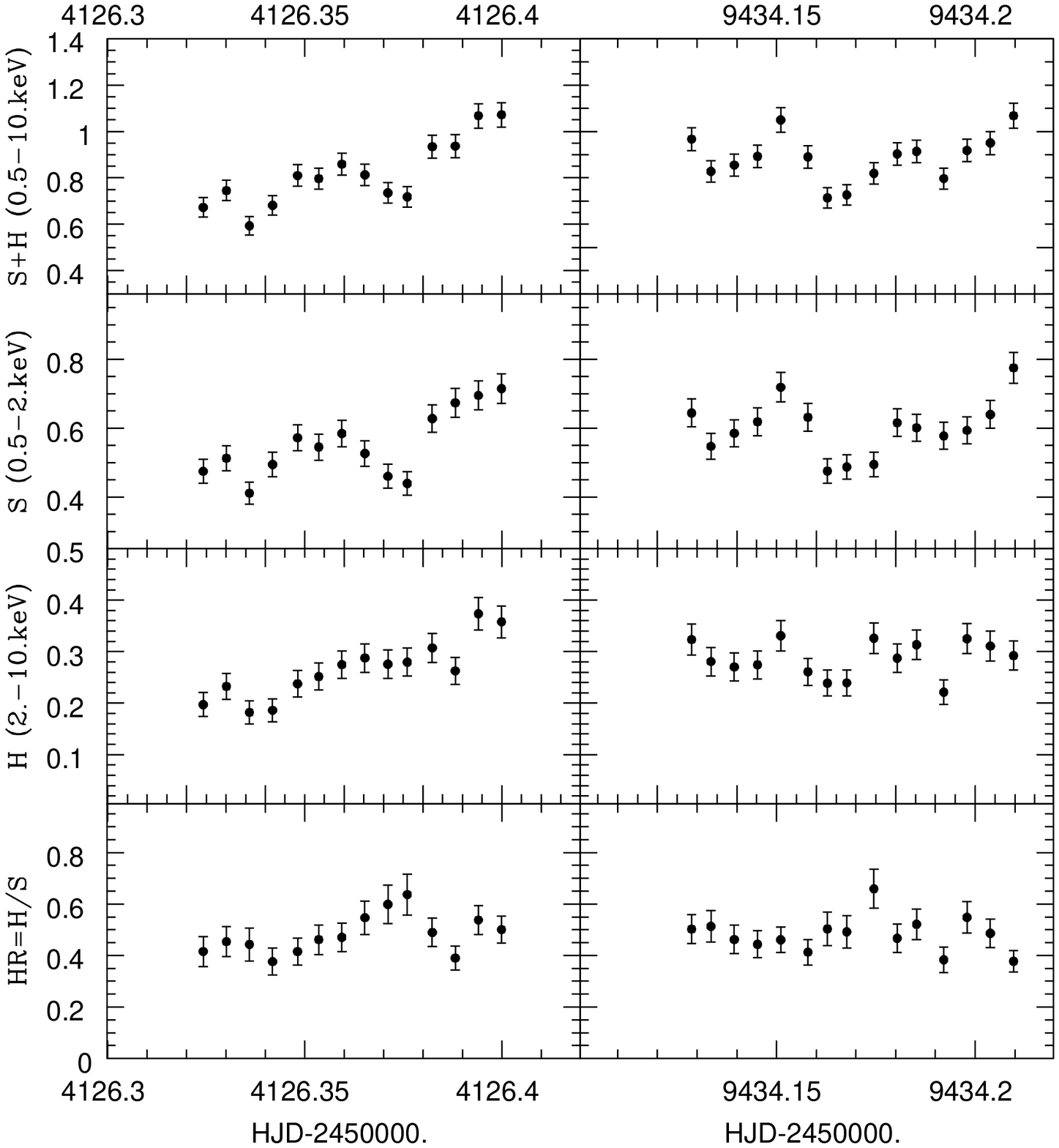}
    \includegraphics[width=5.8cm]{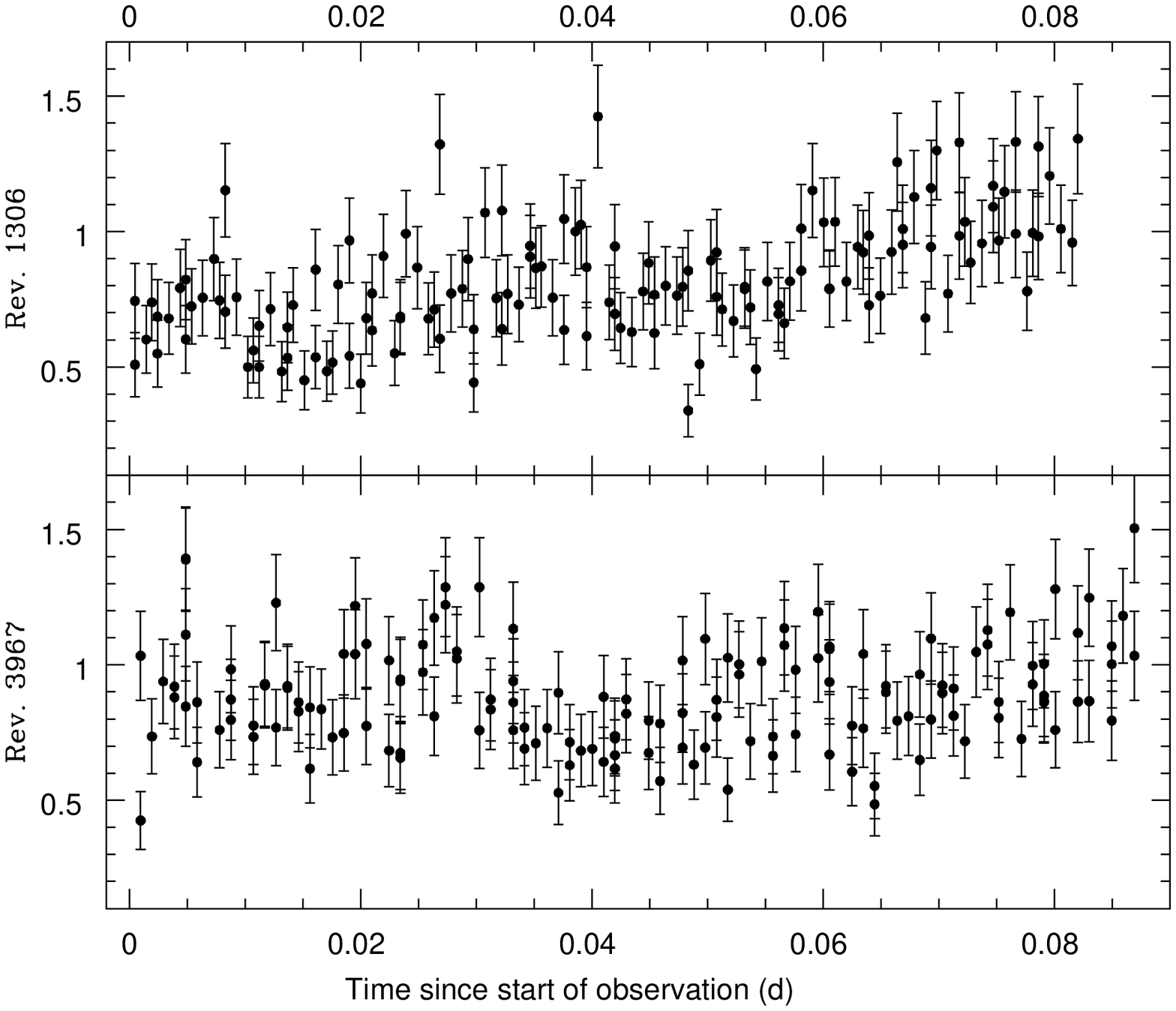}
    \includegraphics[width=5.8cm]{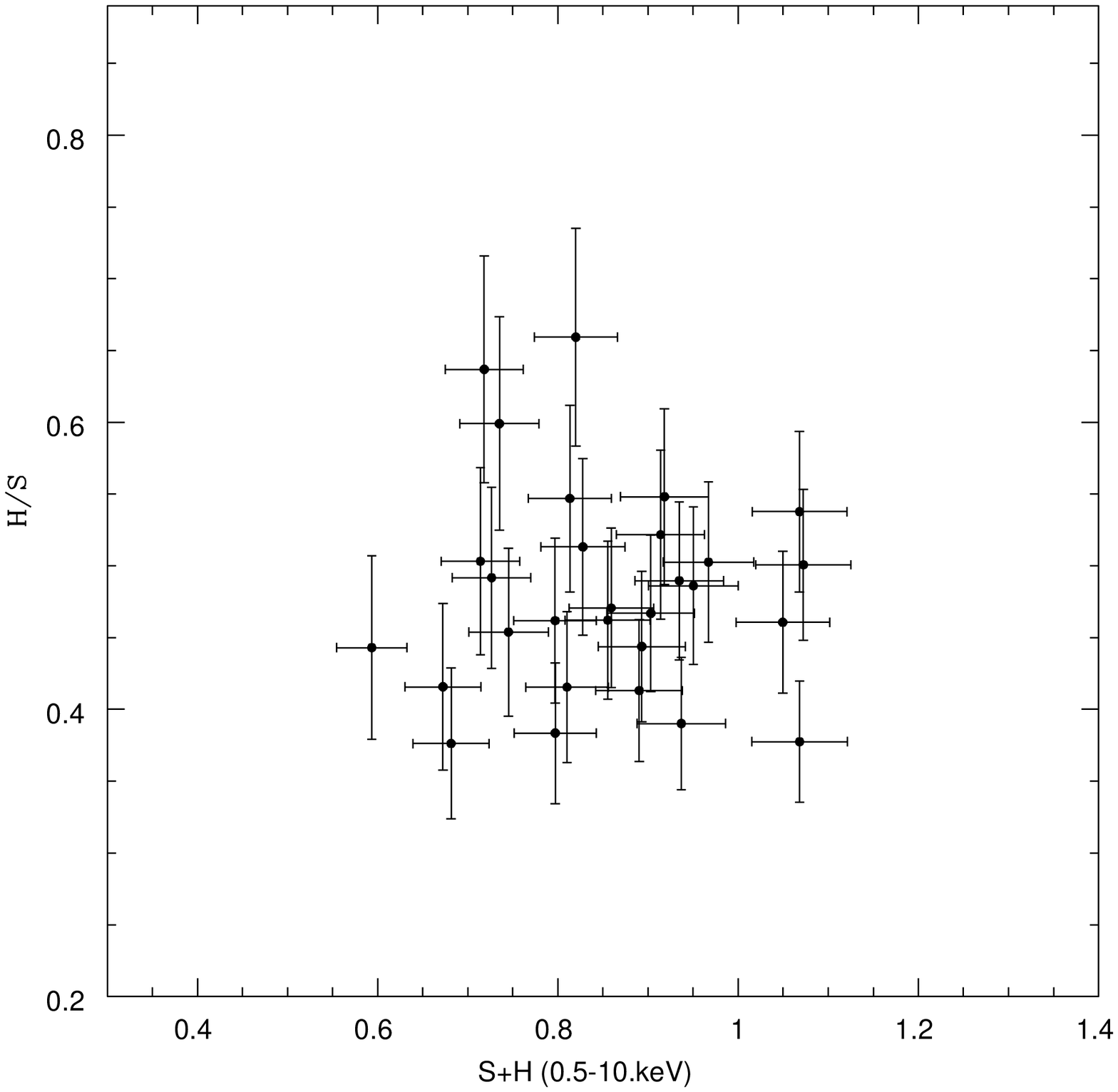}
  \end{center}
  \caption{{\it Left panels:} \xmm\ lightcurves of \vc\ in several energy bands and with a time bin of 500\,s.  {\it Middle panels:} \xmm\ lightcurves of \vc\ in the total energy band and with a time bin of 50\,s. {\it Right panel:} Hardness ratio as a function of the total count rate in all \xmm\ lightcurves of \vc\ with a time bin of 500\,s. }
\label{vclcx}
\end{figure*}

\bsp	
\label{lastpage}
\end{document}